\documentclass[aps,prl,preprint]{revtex4}
\usepackage{epsfig}

\begin{document}

\title{Maxwell's demon through the looking glass}
\author{ Z.~K.~Silagadze} 
\affiliation{Budker Institute of Nuclear Physics and Novosibirsk State 
University, 630 090, Novosibirsk, Russia }

\begin{abstract}
Mechanical Maxwell's demons, such as Smoluchowski's trapdoor and Feynman's
ratchet and pawl need external energy source to operate. If you cease to
feed a demon the Second Law of thermodynamics will quickly stop its operation.
Nevertheless, if the parity is an unbroken symmetry of nature, it may happen
that a small modification leads to demons which do not need feeding. Such
demons can act like perpetuum mobiles of the second kind: extract heat energy 
from only one reservoir, use it to do work and be isolated from the rest of 
ordinary world. Yet the Second Law is not violated because the demons pay 
their entropy cost in the hidden (mirror) sector of the world by emitting 
mirror photons.
\end{abstract}

\maketitle

\section{Introduction}
``The law that entropy always increases - the second law of thermodynamics -
holds, I think, the supreme position among the laws of nature. If someone
points out that your pet theory of the universe is in disagreement with
Maxwell's equations - then so much the worse for Maxwell's equations. If it
is found to be contradicted by experiments - well, these experimentalists do
bungle things sometimes. But if your theory is found to be against the
second law of thermodynamics I can give you no hope; there is nothing for it
but to collapse in deepest humiliation'' \cite{1}. Following Eddington's this
wise advice, I will not challenge the Second Law in this article. Demonic
devices I consider are by no means malignant to the Second Law and their
purpose is not to oppose it but rather to demonstrate possible subtleties of 
the Law's realization in concrete situations.

This philosophy is much in accord with Maxwell's original intention who 
introduced ``a very observant and neat-fingered being'', the Demon, as a means
to circumscribe the domain of validity of the Second law and in particular
to indicate its statistical nature \cite{2,3,4}. 

Later, however, it became traditional to consider demons as a threat to the 
Second Law to be defied. Modern exorcisms use information theory language and 
it is generally accepted that the Landauer's principle \cite{5,6,7} finally 
killed the demon. 

But, as Maxwell's demon was considered to be killed several times in the 
past and every time it resurrected from dead, we have every reason to suspect 
this last information theory exorcism \cite{3,8,9} and believe the old wisdom 
that ``True demons cannot be killed at all'' \cite{10}.

The problem with all exorcisms can be formulated as a dilemma \cite{8}: either
the whole combined system the demon included forms a canonical thermal system 
or it does not. In the first case the validity of the Second Law is assumed 
from the beginning and therefore the demon is predefined to fail. The exorcism
is sound but not profound, although sometimes it can be enlightening and 
delightful.

In the second case it is not evident at all why anti-entropic behaviour can 
not happen. For example, if Hamiltonian dynamics is abandoned, a pressure 
demon can readily be constructed \cite{11}. 
One needs a new physical postulate with independent justification
to ensure the validity of the Second Law for the combined system. Although
Landauer's principle can be proved in several physical situations \cite{12,13}
its generality is not sufficient to exorcise all conceivable extraordinary
demons. For example, the Landauer's principle, as well as ordinary
Gibbsian thermodynamics, fails in the extreme quantum limit then the 
entanglement is essential and the total entropy is not the sum of partial 
entropies of the subsystems \cite{14}.

According to Boltzmann, ``The second law of thermodynamics can be proved from 
the mechanical theory if one assumes that the present state of the universe, 
or at least that part which surrounds us, started to evolve from an improbable
state and is still in a relatively improbable state. This is a reasonable
assumption to make, since it enables us to explain the facts of experience,
and one should not expect to be able to deduce it from anything more
fundamental'' \cite{15}. But how improbable? Roger Penrose estimates 
\cite{15,16} that the initial state of the universe was absurdly improbable:
one part in $10^{10^{123}}$. In the face of this number, I think, you would
rather welcome Maxwell's demon than want to exorcise it. Of course, recent
approaches to the Low Entropy Past problem \cite{17} do not involve demons,
but who can guarantee that the initial low entropy state was prepared without
them? In any case something more fundamental is clearly required to explain
the puzzle and avoid sheer impossibility of the world we observe around us.

It seems, therefore, that exorcism is not productive and it is better to 
allow the presence of demons. However the notion of demons should be 
understood in broader sense: not only creatures which challenge the Second 
Law, but also the ones which use the Second Law in a clever and rather 
unexpected manner.     

The rest of the paper is organized as follows. In the first sections we
consider some classical Maxwell demons. A rather detailed analysis of them
is presented in the framework of the Langevin and Fokker-Planck equations.
I tried to make the exposition as self-contained as possible, assuming that
readers from the mirror matter community are as unfamiliar with some 
subtleties of the Fokker-Planck equation, like It\^{o}-Stratonovich dilemma,
as was the author months ago.

After demonstrating that the demons really work when the temperature 
difference between their two thermal baths is enforced, we speculate about 
a possibility that this temperature difference can be naturally created
if significant amount of mirror matter is added to one of the thermal
reservoirs.

Mirror matter is a hypothetical form of matter expected to exist if nature
is left-right symmetric in spite of {\bf P} and {\bf CP} violations in weak
interactions. An example from the molecular physics is considered to support
this possibility. We cite some review articles where more detailed and 
traditional exposition of the mirror matter idea, as well as relevant 
references, can be found.

The conclusion just finishes the paper with the cheerful remark that mirror
matter demons might be technologically very useful. Therefore it is 
worthwhile to search mirror matter experimentally. It is remarked that one
experiment which searches invisible decays of orthopositronium in vacuum
is under way.

\section{Smoluchowski's trapdoor}
Maxwell's original demon operates a tiny door between two chambers filled
with gas allowing only fast molecules to pass in one direction, and only slow
ones to pass in opposite direction. Eventually a temperature difference is 
generated between the chambers. The demon can be just some purely mechanical 
device to avoid complications connected with the demon's intellect and
information processing. If this device is assumed to be subject of 
conventional thermodynamics (the sound horn of the dilemma mentioned above)
then the demon can succeed only if it hides entropy somewhere else. A very
clever way how to do this was indicated recently \cite{18}.
\begin{figure}[htb]
\begin{center}
\mbox{\epsfig{figure=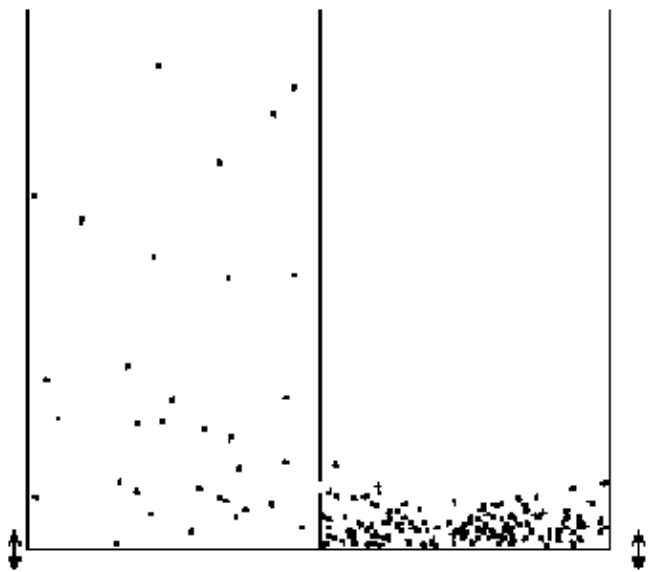}}
\end{center}
\caption {Sand as Maxwell's Demon \cite{18}.}
\label{sand}
\end{figure}

In Fig.\ref{sand} Maxwell's demon left the door  between two chambers
ajar but this shrewd creature had replaced the molecules in the chambers by 
sand grains and set the chambers in vertical vibration by mounting them on a 
shaker. Usually sand grains distribute equally to both sides. But if the 
vibration frequency is lowered below a critical value something remarkable 
happens: the symmetry is spontaneously broken and grains settle 
preferentially on one side. The sand grains act as a Maxwell's demon! This is 
possible because, in contrast to a molecular gas, collisions between grains 
of sand are not elastic and grains are simply heated during the collisions, 
absorbing the entropy \cite{18}.

One of the oldest mechanical Maxwell's demon is the Smoluchowski's trapdoor
\cite{19}. Two chambers filled with gas are connected by an opening that is 
covered by a spring-loaded trapdoor. Molecules from one side tend to push the 
door open and those on the other side tend to push it closed. Naively one 
expects that this asymmetry in the construction of the door turns it into 
a one-way valve which creates a pressure difference between the chambers.

Smoluchowski argued that the thermal fluctuations of the door will spoil the
naive picture and prohibit it to act as a one-way valve, although he did not
provide detailed calculations to prove this fact conclusively. It is 
difficult to trace the trapdoor demon analytically to see how it fails,
but the corresponding computer simulations can be done and the results show
that Smoluchowski was right \cite{19,20}.

The simulations show that the pressure difference between the chambers is
exactly what is expected at equilibrium after the trapdoor's  finite volume 
is taken into account. Therefore the trapdoor can not act as a pressure 
demon.

But if the trapdoor is cooled to reduce its thermal fluctuations the demon 
becomes successful and indeed creates a measurable pressure difference between
the chambers. In simulation program the effective cooling of the door is 
maintained by periodically removing energy from the door then  it is near the 
central partition. As a result the door remains near the closed position 
longer than would occur without the cooling and it indeed operates as a 
one-way valve \cite{19,20}.

The entropy decrease due to this pressure demon is counterbalanced by the 
entropy increase in the refrigerator that cools the door below the ambient gas
temperature. However the demonstration of this fact in computer simulations 
involves some subtleties \cite{20}.

\section{Feynman's ratchet and pawl}
\begin{figure}[htb]
\begin{center}
\mbox{\epsfig{figure=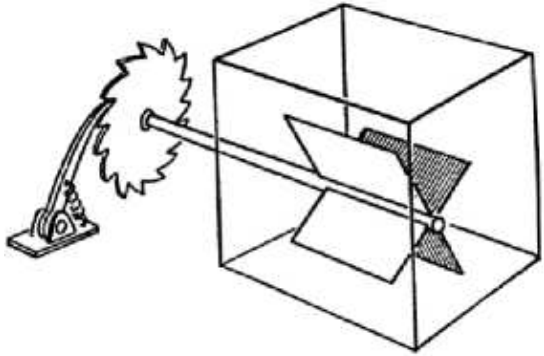}}
\end{center}
\caption {Feynman's ratchet and pawl engine \cite{21}.}
\label{Feyn}
\end{figure}

Close in spirit to the Smoluchowski's trapdoor is Feynman's ratchet and pawl
mechanism \cite{21}. The device (Fig.\ref{Feyn}) contains a box with some gas
and an axle with vanes in it. At the other end of the axle,
outside the box, there is toothed wheel and a pawl that pushes on the cogwheel
through a spring.

If the paddle wheel is small enough, collisions of gas molecules on the vanes
will lead to Brownian fluctuations of the axle. But the ratchet on the other
end side of the axle allows rotation only in one direction and blocks the
counter-rotation. Therefore one expects Feynman's ratchet and pawl engine
to rectify the thermal fluctuations in a similar manner as the use of ratchets
allow windmills to extract useful work from random winds. 

At closer look, however, we realize that the miniature pawl would itself be 
subject to thermal fluctuations that invalidate its rectification property.
In isothermal conditions, when the vanes have the same temperature as the 
ratchet and pawl, the Second Law ensures that the rate of rotations in a wrong
direction due to the pawl's thermal bounces just compensates the rate of the
favorable rotations so that the wheel does a lot of jiggling but no net 
turning. However if the temperatures are different the rate balance no more 
holds and the ratchet and pawl engine really begins to rotate 
\cite{21,22,23,24}.  

Yet there are some subtleties and by way of proof it is better to go into some
details by modeling the ratchet and pawl machine by the following Langevin
equations \cite{25}:
\begin{eqnarray} &&
\lambda\frac{d\theta_1}{dt}=F(\theta_1)-k(\theta_1-\theta_2)+f_1(t), 
\nonumber \\ && \lambda\frac{d\theta_2}{dt}=k(\theta_1-\theta_2)+f_2(t),
\label{Langevin} \end{eqnarray}
where $\theta_1$ and $\theta_2$ are the angular positions of the ratchet and 
the windmill respectively, $\lambda$ is the friction coefficient assumed to
be the same for both baths, 
$F(\theta_1)=-\frac{\partial U}{\partial \theta_1}$ is the torque the ratchet
experiences due to an asymmetric potential $U(\theta_1)$ of its interaction
with the pawl. It is assumed that the ratchet and the windmill are connected
by an elastic spring of rigidity $k$. Finally, $f_1(t)$ and $f_2(t)$ represent
Brownian random torques on the ratchet and on the windmill respectively due to
thermal fluctuations. It is assumed that these random torques constitute two 
independent Gaussian white noises
\begin{equation}
<f_i(t)>=0,\;\;\;<f_i(t)\,f_j(t^\prime)>=2k_BT_i\lambda\,\delta_{ij}\,
\delta(t-t^\prime),
\label{wnoise} \end{equation}
$k_B$ being the Boltzmann constant. These Langevin equations correspond to 
the over-dumped regime (with neglected inertia).

It is more convenient to introduce another set of angular variables
\begin{equation}
\Theta=\frac{1}{2}(\theta_1+\theta_2),\;\;\; \theta=\frac{1}{2}
(\theta_1-\theta_2).
\label{Theta} \end{equation}
It is the variable $\Theta$ which describes the net rotation of the system
and is, therefore, the only object of our interest, while the variable
$\theta$ describes the relative motion and is irrelevant for our goals. In
terms of these variables, equations (\ref{Langevin}) read
\begin{eqnarray} &&
\lambda\,\dot{\Theta}=\frac{1}{2}F(\Theta+\theta)+
\frac{1}{2}\left [ f_1(t)+f_2(t) \right ], 
\nonumber \\ && \lambda\,\dot{\theta}=\frac{1}{2}F(\Theta+\theta)-
2k\theta+\frac{1}{2}\left [ f_1(t)-f_2(t) \right ].
\label{Langevin1} \end{eqnarray}
The relative rotation $\theta$ arises due to the Brownian jiggling of the 
ratchet and of the windmill and hence is expected to be very small. 
Therefore one can expand 
$$F(\Theta+\theta)\approx F(\Theta)+\theta\,F^{\,\prime}(\Theta), \;\;
F^{\,\prime}(\Theta)\equiv \frac{dF}{d\Theta}.$$
Besides, the dynamics of the variable $\theta$ is very rapid and at any time 
scale, relevant for the evolution of the slow variable $\Theta$, the fast
variable $\theta$ will be able to relax to a quasi-stationary value given by
setting $\dot{\theta}=0$ in the second equation of (\ref{Langevin1}):
\begin{equation}
\theta\approx\frac{F(\Theta)+f_1(t)-f_2(t)}{4k-F^{\,\prime}(\Theta)}.
\label{stheta} \end{equation}
This allows to eliminate $\theta$ from (\ref{Langevin1}) and arrive at the 
following equation for the relevant variable $\Theta$ \cite{25}:
\begin{equation}
\dot{\Theta}=H(\Theta)+g_1(\Theta)f_1(t)+g_2(\Theta)f_2(t),
\label{Ltheta} \end{equation} 
where
$$H(\Theta)=\frac{F(\Theta)}{2\lambda}\left [ 1+\frac{F^{\,\prime}(\Theta)}
{4k}\right],\;\;g_1(\Theta)=\frac{1}{2\lambda}\left [ 1+\frac{F^{\,\prime}
(\Theta)}{4k}\right],\;\;g_2(\Theta)=\frac{1}{2\lambda}\left [ 1-\frac{F^{\,
\prime}(\Theta)}{4k}\right],$$
and terms of the second and higher order in $\frac{F^{\,\prime}(\Theta)}
{4k}$ were neglected.

The resulting Langevin equation (\ref{Ltheta}) is the stochastic differential
equation with multiplicative noises and, therefore, subject to the notorious
It\^{o}-Stratonovich dilemma \cite{26,27}. The problem is that stochastic 
integrals one needs to calculate various average values are in general 
ill-defined without additional interpretation rules (the white noise is a 
singular object after all, like the Dirac's $\delta$-function). The most 
commonly used It\^{o} and Stratonovich interpretations of stochastic integrals
lead to different results when multiplicative noise is present. 

From the physics side, this ill-definedness may be understood as follows 
\cite{27}. According to (\ref{Ltheta}) each $\delta$-pulse in $f_1(t)$ or 
$f_2(t)$ gives rise to a pulse in $\dot {\Theta}$ and hence an instantaneous 
jump in $\Theta$. Then it is not clear which value of $\Theta$ should be used 
in $g_1(\Theta)$ and $g_2(\Theta)$: the value just before the jump, after the 
jump, or some average of these two. It\^{o} prescription assumes that the 
value before the jump should be used, while Stratonovich prescription 
advocates for the mean value between before and after the jump.

Our manipulations, which led to (\ref{Ltheta}) from (\ref{Langevin}), already
assumes the Stratonovich interpretation because we had transformed variables 
in (\ref{Langevin}) as if  (\ref{Langevin}) were ordinary (not stochastic)
differential equations and this is only valid for the Stratonovich 
interpretation \cite{27}.

In fact there is a subtle point here. The naive adiabatic elimination of the
fast variable we applied for $\theta$ (by setting $\dot{\theta}=0$) not 
necessarily implies the Stratonovich interpretation \cite{28}. Depending on 
the fine details of the physical system and of the limiting process, one may
end with Stratonovich, It\^{o} or even with some other interpretation which is 
neither It\^{o} nor Stratonovich \cite{28}. Nevertheless the Stratonovich 
interpretation will be assumed in the following like as was done tacitly in 
\cite{25}.
 
Let $P(\Theta,t)$ be the probability density for the stochastic process 
$\Theta(t)$. Then
\begin{equation}
P(\Theta,t)=\int\limits_{-\infty}^\infty G(\Theta,t;\,\Theta_0,t_0)\,
P(\Theta_0,t_0)\,d\Theta_0,\;\;t>t_0,
\label{Pdens} \end{equation}
where the Green's function (the conditional probability density over $\Theta$ 
at time $t$ under the condition that $\Theta=\Theta_0$ at $t=t_0$) satisfies 
the initial value equation
\begin{equation}
G(\Theta,t_0;\,\Theta_0,t_0)=\delta(\Theta-\Theta_0).
\label{Icond} \end{equation}
Now consider
\begin{equation}
\frac{\partial P(\Theta,t)}{\partial t}=\lim_{\Delta t\to 0}
\frac{P(\Theta,t+\Delta t)-P(\Theta,t)}{\Delta t}.
\label{dPdt} \end{equation}
According to (\ref{Pdens})
\begin{equation}
P(\Theta,t+\Delta t)=\int\limits_{-\infty}^\infty G(\Theta,t+\Delta t;\,
\Theta_0,t)\,P(\Theta_0,t)\,d\Theta_0.
\label{DtPdens} \end{equation}
As the time interval $\Delta t$ is very short, the function $G(\Theta,t+
\Delta t;\,\Theta_0,t)$ can differ from its initial $\delta$-function value
only slightly by drifting and broadening a little which can be modeled by the
drift coefficient $D^{(1)}(\Theta)$ and the diffusion coefficient 
$D^{(2)}(\Theta)$ respectively if we expand \cite{29}
\begin{equation}
G(\Theta,t+\Delta t;\,\Theta_0,t)\approx \delta(\Theta-\Theta_0)+
D^{(1)}(\Theta_0)\,\Delta t \,\delta^{(1)}(\Theta-\Theta_0)+
D^{(2)}(\Theta_0)\,\Delta t \,\delta^{(2)}(\Theta-\Theta_0),
\label{dexpan} \end{equation}
where $\delta^{(n)}(\Theta)$ denotes the $n$-th order derivative of the
$\delta$-function with the basic property
$$\int\limits_{-\infty}^\infty f(\Theta)\,\delta^{(n)}(\Theta-\Theta_0)\,
d\Theta=\frac{(-1)^{\,n}}{n\,!}\,\left . \frac{d^{\,n}f(\Theta)}
{d\Theta^n}\right |_{\Theta=\Theta_0}. $$ 
Substituting (\ref{dexpan}) into (\ref{DtPdens}) we get
\begin{equation}
P(\Theta,t+\Delta t)\approx P(\Theta,t)+\Delta t\,\frac{\partial}
{\partial \Theta}\left [D^{(1)}(\Theta)\,P(\Theta,t)\right ]+\frac{\Delta t}
{2}\,\frac{\partial^2}{\partial \Theta^2}\left [D^{(2)}(\Theta)\,
P(\Theta,t)\right ]
\label{Pdt} \end{equation}
and therefore (\ref{dPdt}) implies the following Fokker-Planck equation
\begin{equation}      
\frac{\partial P(\Theta,t)}{\partial t}=\frac{\partial}{\partial \Theta}
\left [D^{(1)}(\Theta)\,P(\Theta,t)\right ]+\frac{1}{2}\,\frac{\partial^2}
{\partial \Theta^2}\left [D^{(2)}(\Theta)\, P(\Theta,t)\right ].
\label{FP} \end{equation}
The Fokker-Planck equation determines the evolution of the probability density
provided the drift and diffusion coefficient functions are known. These
functions are related to the first two moments the initially localized 
density function develops in the short time interval $\Delta t$ because
(\ref{Pdt}) indicates that
\begin{eqnarray} &&
<\Delta \Theta>\,=\int\limits_{-\infty}^\infty (\Theta-\Theta_0)\, 
P(\Theta,t+\Delta t)\,d\Theta=-D^{(1)}(\Theta_0)\,\Delta t, \nonumber \\ &&
<(\Delta \Theta)^2>\,=\int\limits_{-\infty}^\infty (\Theta-\Theta_0)^2\, 
P(\Theta,t+\Delta t)\,d\Theta=D^{(2)}(\Theta_0)\,\Delta t,
\label{D12} \end{eqnarray}
if $P(\Theta,t)=\delta(\Theta-\Theta_0)$. 

On the other hand, these moments can be calculated directly from the Langevin
equation. Integrating (\ref{Ltheta}), we get
$$\Delta \Theta=\int\limits_t^{t+\Delta t} H(\Theta)\,dt^\prime+
\int\limits_t^{t+\Delta t} g_1(\Theta)f_1(t^\prime)\,dt^\prime+
\int\limits_t^{t+\Delta t} g_2(\Theta)f_2(t^\prime)\,dt^\prime.$$
According to the Stratonovich prescription, in the last two stochastic 
integrals $\Theta$ should be replaced with
$$\frac{\Theta(t)+\Theta(t+\Delta t)}{2}\approx\Theta(t)+\frac{1}{2}
\dot{\Theta}(t)\,\Delta t=\Theta_0+\frac{\Delta t}{2}\left [ H(\Theta_0)+
g_1(\Theta_0)f_1(t)+g_2(\Theta_0)f_2(t)\right ],$$
where $\Theta_0=\Theta(t)$. But
$$g(\Theta_0+\Delta\Theta)\approx g(\Theta_0)+\Delta \Theta \left .
\frac{\partial g}{\partial \Theta}\right |_{\Theta=\Theta_0}.$$
Therefore we obtain
$$\Delta \Theta\approx H(\Theta_0)\,\Delta t +
\frac{\Delta t}{2}\left [ H(\Theta_0)+g_1(\Theta_0)f_1(t)+g_2(\Theta_0)f_2(t)
\right ]\times $$
\begin{equation}
\left [\frac{\partial g_1}{\partial \Theta}(\Theta_0)
\int\limits_t^{t+\Delta t} f_1(t^\prime)\,dt^\prime+
\frac{\partial g_2}{\partial \Theta}(\Theta_0)\int\limits_t^{t+\Delta t} 
f_2(t^\prime)\,dt^\prime \right ]+
g_1(\Theta_0)\int\limits_t^{t+\Delta t} f_1(t^\prime)\,dt^\prime+
g_2(\Theta_0)\int\limits_t^{t+\Delta t} f_2(t^\prime)\,dt^\prime. 
\label{Dtheta} \end{equation}
Taking an ensemble average by using (\ref{wnoise}), we get
$$<\Delta \Theta>\,=\left [H(\Theta_0)+k_BT_1\lambda\,g_1(\Theta_0)\,
\frac{\partial g_1}{\partial \Theta}(\Theta_0)+k_BT_2\lambda\,
g_2(\Theta_0)\,\frac{\partial g_2}{\partial \Theta}(\Theta_0)\right ]\,
\Delta t.$$
While averaging the square of (\ref{Dtheta}) gives
$$<(\Delta \Theta)^2>\,=2k_B\lambda\left [T_1\,g_1^2(\Theta_0)+
T_2\,g_2^2(\Theta_0)\right ]\Delta t, $$
because
$$\int\limits_t^{t+\Delta t}dt^\prime
\int\limits_t^{t+\Delta t}dt^{\prime\prime}<f_i(t^\prime)f_j(t^{\prime
\prime})>\,=2k_BT_i\lambda\,\delta_{ij}\,\Delta t.$$
Comparing these expressions with (\ref{D12}), we see that
\begin{equation}
D^{(1)}(\Theta)=-H(\Theta)-k_BT_1\lambda \,g_1(\Theta)\,
\frac{\partial g_1}{\partial \Theta}-
k_BT_2\lambda \,g_2(\Theta)\,\frac{\partial g_2}{\partial \Theta},
\label{D1}\end{equation}
and
\begin{equation}
D^{(2)}(\Theta)=2k_B\lambda\left [T_1\,g_1^2(\Theta)+T_2\,g_2^2(\Theta)
\right ].
\label{D2}\end{equation}
If we substitute (\ref{D1}) and (\ref{D2}) into the Fokker-Planck equation
(\ref{FP}), it takes the form
\begin{equation}
\frac{\partial P(\Theta,t)}{\partial t}+\frac{\partial J(\Theta,t)}
{\partial \Theta}=0,
\label{FPJ} \end{equation}
where the probability current
\begin{equation}
J(\Theta,t)=H(\Theta)P(\Theta,t)-k_BT_1\lambda\,g_1(\Theta)\,
\frac{\partial}{\partial \Theta}\left [ g_1(\Theta)P(\Theta,t)\right ]-
k_BT_2\lambda\,g_2(\Theta)\,\frac{\partial}{\partial \Theta}\left [ 
g_2(\Theta)P(\Theta,t)\right ].
\label{J} \end{equation}
It can easily be checked that \cite{25}
\begin{equation}
J(\Theta,t)=H(\Theta)P(\Theta,t)-g(\Theta)\,
\frac{\partial}{\partial \Theta}\left [ g(\Theta)P(\Theta,t)\right ],
\label{Jg} \end{equation}
with 
\begin{equation}
g(\Theta)=\sqrt{k_BT_1\lambda\,g_1^2(\Theta)+k_BT_2\lambda\,g_2^2
(\Theta)}.
\label{gtheta} \end{equation}
We are interested in a steady state operation of the engine $P(\Theta,t)=
P(\Theta)$. Then the Fokker-Planck equation (\ref{FPJ}) and the relation 
(\ref{J}) show that the probability current depends neither on time nor
angular position: $J(\Theta,t)=J_0$ is a constant. From (\ref{Jg}) we get
the following differential equation
\begin{equation}
\frac{dP(\Theta)}{d\Theta}+\frac{1}{g(\Theta)}\frac{dg(\Theta)}{d\Theta}\,
P(\Theta)-\frac{H(\Theta)}{g^2(\Theta)}\,P(\Theta)=-\frac{J_0}{g^2(\Theta)}.
\label{DEP} \end{equation}

The equation (\ref{DEP}) is a linear differential equation and can be solved
in a standard way. The solution is \cite{30}
\begin{equation}
P(\Theta)=\left [A_0-J_0\int\limits_0^\Theta \frac{e^{-B(\Theta^\prime)}}
{g(\Theta^\prime)}\,d\Theta^\prime \right ] \frac{e^{B(\Theta)}}{g(\Theta)},
\label{P} \end{equation}
where $A_0$ is some constant and
$$B(\Theta)=\int\limits_0^\Theta \frac{H(\Theta^\prime)}{g^2(\Theta^\prime)}
\,d\Theta^\prime .$$
The periodic boundary conditions $P(0)=P(2\pi),\, g(0)=g(2\pi)$ imply that 
the constants $J_0$ and $A_0$ are interconnected:
$$A_0=J_0\,\frac{e^\beta}{e^\beta-1}\int\limits_0^{2\pi}\frac{e^{-B(\Theta)}}
{g(\Theta)}\,d\Theta,$$
where
\begin{equation}
\beta=B(2\pi)=\int\limits_0^{2\pi} \frac{H(\Theta)}{g^2(\Theta)}
\,d\Theta .
\label{beta} \end{equation}
After a little algebra, we find that
\begin{equation}
P(\Theta)=J_0\,\frac{e^{B(\Theta)}}{g(\Theta)}\left [ \; \int\limits_\Theta^
{2\pi} \frac{e^{-B(\Theta^\prime)}}{g(\Theta^\prime)}\,d\Theta^\prime+\frac{1}
{e^\beta-1}\int\limits_0^{2\pi}\frac{e^{-B(\Theta^\prime)}}{g(\Theta^\prime)}
\,d\Theta^\prime \right ].
\label{Ptheta} \end{equation} 
The normalization condition
$$\int\limits_0^{2\pi}P(\Theta)\,d\Theta=1$$
then determines the probability current $J_0$ to be
\begin{equation}
J_0=\frac{e^\beta-1}{I},
\label{J0} \end{equation}
where
$$I=(e^\beta-1)\int\limits_0^{2\pi}\frac{e^{B(\Theta)}}{g(\Theta)}
\,d\Theta \int\limits_\Theta^{2\pi}\frac{e^{-B(\Theta^\prime)}}{g(\Theta^
\prime)}\,d\Theta^\prime+ \int\limits_0^{2\pi}\frac{e^{B(\Theta)}}{g(\Theta)}
\,d\Theta \int\limits_0^{2\pi}\frac{e^{-B(\Theta^\prime)}}{g(\Theta^\prime)}
\,d\Theta^\prime.$$
The angular velocity $\omega(\Theta)$ can be determined from the relation
$$J(\Theta)=P(\Theta)\,\omega(\Theta).$$
Its average (the net angular velocity of the ratchet and pawl engine) equals 
to
\begin{equation}
\omega=\int\limits_0^{2\pi}\omega(\Theta)\,P(\Theta)\,d\Theta=
\int\limits_0^{2\pi} J(\Theta)\,d\Theta=2\pi\,J_0.
\label{omega} \end{equation}
As we see from (\ref{omega}) and (\ref{J0}), there is no net angular velocity
if $\beta=0$. But from (\ref{beta})
\begin{equation}
\beta\approx 2\int\limits_0^{2\pi}\frac{F(\Theta)}{k_BT_1}\left [1+
\frac{F^\prime(\Theta)}{4k}\right ]\,\frac{1}{1+\frac{T_2}{T_1}+\left (
1-\frac{T_2}{T_1}\right )\frac{F^\prime(\Theta)}{2k}}\,d\Theta.
\label{betaT} \end{equation}
Note that 
$$\int\limits_0^{2\pi}F(\Theta)\,d\Theta=-\left [U(2\pi)-U(0)\right ]=0
\;\;\;\mbox{and}\;\;\;
\int\limits_0^{2\pi}F(\Theta)F^\prime(\Theta)\,d\Theta=\frac{1}{2}\left [
F^2(2\pi)-F^2(0)\right ]=0,$$
because of the periodic boundary conditions. Therefore, if $T_1=T_2$, then
$\beta=0$ and there is no net angular velocity, as is demanded by the Second
Law. Less trivial fact, which follows from (\ref{betaT}), is that for 
absolutely rigid axle, $k\to\infty$, the engine does not work either.

\section{Brillouin's demon}
Although Feynman's ratchet and pawl gadget is fascinating, its experimental
realization is doubtful because it does not seem feasible to arrange the 
necessary temperature gradient at nanoscale without generating violent
convective currents in ambient material \cite{23}. In this respect its 
electrical counterpart, the Brillouin diode demon is more promising.

The Brillouin diode engine \cite{31,32} consists of a diode in parallel to
a resistor and a capacitor. Thermal noise in the resistor makes it the AC
voltage source. Naively, one expects the diode to rectify this AC voltage and
the capacitor to become self-charged. However the Second Law prohibits such
a demon devise to operate unless the diode and the resistor are kept at 
different temperatures -- in complete analogy with the ratchet and pawl
engine.

Both the Feynman ratchet and pawl gadget \cite{22,23} and the diode engine 
\cite{33} are very inefficient thermal engines with efficiencies far below
the Carnot value. The reason is that they operate irreversibly due to a
unavoidable heat exchange between the two thermal reservoirs the engine is
simultaneously in contact. In the ratchet and pawl case, it is the mechanical
coupling between the vanes and the ratchet  which induces, via fluctuations, 
a heat transfer between the reservoirs, even under negligible thermal 
conductivity of the axle \cite{22}.  

\begin{figure}[htb]
\begin{center}
\mbox{\epsfig{figure=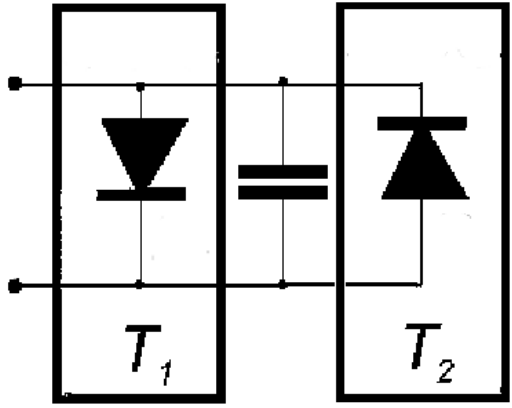,height=6cm}}
\end{center}
\caption {Brillouin's diode engine \cite{34}.}
\label{RCD}
\end{figure}

It was shown in \cite{34} that the heat exchange between the reservoirs of
the diode engine is significantly reduced if the resistor in the circuit is 
replaced by the second diode switched in the opposite direction as shown in
Fig.\ref{RCD}. Let us analyze this system considering diodes as just nonlinear
resistors \cite{34}.

If $u$ is the voltage of the capacitor and $i_1,i_2$ are currents through two
nonlinear resistors then
\begin{equation}
i_1R_1(u)+u=v_1(t)\;\;\; \mbox{and} \;\;\; i_2R_2(u)+u=v_2(t),
\label{i12} \end{equation}
where $v_1(t)$ and $v_2(t)$ are Nyquist stochastic electromotive forces 
\cite{35}, due to thermal agitation in the corresponding resistors, 
satisfying \cite{36}
\begin{equation}
<v_i(t)>\,=0,\;\;\;<v_i(t)\,v_j(t^\prime)>\,=2k_BT_iR_i\,\delta_{ij}\,
\delta(t-t^\prime).
\label{Nuquist} \end{equation}
But $i_1+i_2=\dot{q}$, where $q=Cu$ is the charge of the capacitor with 
capacitance $C$. Therefore (\ref{i12}) indicate the following Langevin 
equation
\begin{equation}
\dot{u}=-\frac{u}{R(u)\,C}+\frac{v_1(t)}{R_1(u)\,C}+\frac{v_2(t)}{R_2(u)\,C},
\label{Ldiode} \end{equation}
where
$$R(u)=\frac{R_1(u)\,R_2(u)}{R_1(u)+R_2(u)}.$$
Using Stratonovich prescription, the drift and diffusion coefficient 
functions can be calculated from (\ref{Ldiode}) in the manner described in 
the previous section for the equation (\ref{Ltheta}). the results are
\begin{equation}
D^{(1)}(u)=\frac{u}{R(u)\,C}+\frac{k_BT_1R_1^\prime(u)}{R_1^2(u)\,C^2}+
\frac{k_BT_2R_2^\prime(u)}{R_2^2(u)\,C^2},\;\;\;
D^{(2)}(u)=\frac{2k_B}{C^2}\left [\frac{T_1}{R_1(u)}+\frac{T_2}{R_2(u)}
\right ],
\label{D12diode}\end{equation} 
where the prime denotes differentiation with respect to $u$.

The Fokker-Planck equation that follows
$$\frac{\partial P(u,t)}{\partial t}+\frac{\partial J(u,t)}{\partial u}=0$$
has the following probability current
\begin{equation}
J(u,t)=-\frac{uP(u,t)}{R(u)\,C}-\frac{k_B}{C^2}\left (\frac{T_1}{R_1(u)}+
\frac{T_2}{R_2(u)}\right )\,\frac{\partial P(u,t)}{\partial u}
\label{FPdiode} \end{equation}
in agreement with \cite{33,34}.

The steady state operation (stationary solution of (\ref{FPdiode})) now 
corresponds to the vanishing probability current because evidently 
$J(\pm\infty,t)=0$. Then (\ref{FPdiode}) produces a simple homogeneous linear
differential equation for $P(u)$ with the solution 
\begin{equation}
P(u)=P_0\,\exp{\left \{-\frac{C}{k_B}\int\limits_0^u\frac{R_1(u)+R_2(u)}
{T_1R_2(u)+T_2R_1(u)}\,u\,du\right \} }.
\label{Pu} \end{equation}
If $T_1=T_2=T$, (\ref{Pu}) reduces to the Boltzmann distribution
$$P(u)=P_0\,\exp{\left \{-\frac{Cu^2}{2k_BT}\right \}}$$
as should be expected for the capacitor's energy in thermal equilibrium.
The Boltzmann distribution is symmetric in $u$ and therefore $<u>\,=0$.
Not surprisingly, the Second Law wins in isothermal situation irrespective
of the volt-ampere characteristics of the diodes, and no self-charging of 
the capacitor takes place.

If temperatures are different, however, the distribution (\ref{Pu}) is no
more symmetric (note that for identical diodes $R_2(u)=R_1(-u)$). Therefore,
if you feed the Brillouin demon with an external energy input, which 
maintains the necessary temperature difference, it will operate successfully.

Let us now consider the heat exchange between the two thermal baths 
\cite{36,37}. During a small time period $\Delta t$, the electromotive force
$v_1$ performs a work $\int i_1v_1dt$ and therefore this amount of energy is
taken from the first thermal bath at temperature $T_1$. But a part of this
energy, namely $\int i_1^2 R_1dt$, is returned back to the bath as the 
nonlinear resistor dissipates the Joule heat into the bath. Therefore, the 
net energy taken from the first bath equals
$$\Delta Q_1=\int\limits_t^{t+\Delta t}i_1(v_1-i_1R_1)\,dt^\prime=
\int\limits_t^{t+\Delta t}i_1u\,dt^\prime=\int\limits_t^{t+\Delta t}
\frac{v_1-u}{R_1}\,u\,dt^\prime\approx-\frac{u^2}{R_1}\Delta t+
\int\limits_t^{t+\Delta t}\frac{u}{R_1}\,v_1(t^\prime)\,dt^\prime.$$
In the remaining stochastic integral, $\frac{u}{R_1}$ has to be replaced,
according to the Stratonovich prescription, with its value at the middle of
the time interval, at time $t+\frac{\Delta t}{2}$, which approximately equals
$$\frac{u}{R_1}+\frac{\Delta t}{2}\,\frac{d}{du}\left (\frac{u}{R_1}\right )
\dot {u}=\frac{u}{R_1}+\frac{\Delta t}{2}\,\frac{d}{du}\left (\frac{u}{R_1}
\right )\left (-\frac{u}{R\,C}+\frac{v_1(t)}{R_1\,C}+\frac{v_2(t)}{R_2\,C}
\right ), $$ 
where now $u$ is evaluated at time $t$. Therefore, taking an ensemble average,
we get
$$<\Delta Q_1>\,=\left [ -\frac{u^2}{R_1}+\frac{k_BT_1}{C}\,\frac{d}{du}
\left (\frac{u}{R_1}\right )\,\right ]\Delta t.$$
Now we average $<\Delta Q_1>/\Delta t$ over the voltage distribution $P(u)$
and get the heat absorbed from the first reservoir at temperature $T_1$ per
unit time 
$$\dot{Q}_1=\int\limits_{-\infty}^\infty \left [ -\frac{u^2}{R_1}+
\frac{k_BT_1}{C}\,\frac{d}{du}\left (\frac{u}{R_1}\right )\,\right ]P(u)\,du
=-\int\limits_{-\infty}^\infty \left [\frac{k_BT_1}{CR_1(u)}\,\frac{\partial
P(u)}{\partial u}+\frac{u}{R_1(u)}P(u)\right ] u\,du $$
in agreement with \cite{33,34}. The last step here follows from 
$P(\pm\infty)=0$ when integration by parts is applied.

Note that
$$\dot{Q}_1+\dot{Q}_2=-\int\limits_{-\infty}^\infty \left [ \frac{u}{R(u)}P(u)
+\frac{k_B}{C}\left (\frac{T_1}{R_1(u)}+\frac{T_2}{R_2(u)}\right )
\frac{\partial P(u)}{\partial u}\right ] u\,du=\int\limits_{-\infty}^\infty
J(u)\,u\,du=0$$
when J=0. In other words, the heat dissipated into the second reservoir at 
temperature $T_2$ per unit time equals to the heat absorbed from the first 
reservoir per unit time. Therefore $\dot{Q}\equiv\dot{Q}_1$ is just the heat 
flux from the first thermal bath to the second and
\begin{equation}
\dot{Q}=\int\limits_{-\infty}^\infty \left [ -\frac{u^2}{R_1(u)}+
\frac{k_BT_1}{C}\,\frac{d}{du}\left (\frac{u}{R_1(u)}\right )\,
\right ]P(u)\,du
\label{Qdot} \end{equation}
To have some impression of the magnitude of this flux, we approximate the
volt-ampere characteristics of the diodes by a step function
$$R_1(u)=R_2(-u)=R_+\,\theta(u)+R_-\,\theta(-u)=\left \{ \begin{tabular}{cc}
$R_+$, if $u>0$, \\ $R_-$, if $u<0$. \end{tabular}
\right . $$
When
$$\frac{d}{du}\left (\frac{u}{R_1(u)}\right )=\frac{d}{du}\left [\frac{u}
{R_+}\theta(u)+\frac{u}{R_-}\theta(-u)\right ]=\frac{\theta(u)}{R_+}+
\frac{\theta(-u)}{R_-}+\left (\frac{1}{R_+}-\frac{1}{R_-}\right )
u\,\delta(u).$$
But $u\,\delta(u)=0$ and a straightforward calculation gives the heat flux
which is linear in the temperature difference \cite{34}
\begin{equation}
\dot{Q}=\frac{k_B(T_1-T_2)}{C(R_++R_-)}.
\label{hflux} \end{equation}
For ideal diodes with infinite backward resistance, $\dot{Q}=0$ and there is
no heat exchange between thermal reservoirs under zero load. Therefore one
expects that the efficiency of the ideal diode engine tends to the Carnot 
efficiency $1-T_2/T_1$ when the external load tends to zero. This fact was 
indeed demonstrated in \cite{34}.

\section{Mirror World and how it can assist demons}
Mirror world was introduced in 1966 by Kobzarev, Okun, and Pomeranchuk 
\cite{38} (a first-hand historical account can be found in \cite{39}),
although the basic idea dates back to Lee and Yang's 1956 paper \cite{40}
and subsequently was rediscovered in the modern context of renormalizable 
gauge theories in \cite{42}.

The idea behind mirror matter is simple and can be explained as follows 
\cite{43}. The naive parity operator {\bf P}, representing the space 
inversion, interchanges left and right. But in case of internal symmetries, 
when there are several equivalent left-handed states and several equivalent 
right-handed states, it is not a priori obvious what right-handed state should 
correspond to a given left-handed state. All operators of the type {\bf SP},
where {\bf S} is an internal symmetry operator, are equivalent. If we can find
some internal symmetry {\bf M}, for which {\bf MP} remains unbroken in the 
real world, then we can say that the world is left-right symmetric in over-all
because {\bf MP} is as good as the parity operator, as is {\bf P} itself.

What remains is to find an appropriate internal symmetry {\bf M}. Not every
choice of {\bf M} leads to the mirror world in a sense of a decoupled hidden 
sector. For example, the most economical first try is the charge conjugation
operator {\bf C} in the role of {\bf M} \cite{44,45,46}. In this case mirror
world coincides with the world of antiparticles. But {\bf CP} is not 
conserved and therefore such a world is not left-right symmetric. 

Then the most natural and secure way to enforce the left-right symmetry is to 
double our Standard Model world by introducing a mirror twin for every 
elementary particle in it and arrange the right-handed mirror weak 
interactions, so that every {\bf P}-asymmetry in ordinary world is accompanied 
by the opposite {\bf P}-asymmetry in the hidden mirror sector. This new mirror 
sector must be necessarily hidden, that is only very weak interactions between
mirror and ordinary particles may be allowed, otherwise this scenario will 
come to immediate conflict with phenomenology \cite{38}.

If the parity symmetry in our world is indeed restored at the expense of the 
hidden mirror sector, many interesting phenomenological and astrophysical
consequences follow that were discussed many times in the literature. It would 
be boring to repeat all this here. Therefore I cite only some review articles 
\cite{39,47,48,49,50} where relevant references can be found. Instead of 
following the well-known trails, I choose a new pathway to the mirror world,
with boron trifluoride ($BF_3$) as our rather exotic guide.  

Boron trifluoride is a highly toxic, colorless, nonflammable gas with a 
pungent odor, used heavily in the semiconductor industry. But what is relevant
for us here is the shape of its planar molecule. Three fluorine atoms seat at 
the corners of a equilateral triangle with the boron atom in the center.
This shape is obviously parity invariant, with parity identified with the 
reflection in the $y$-axis of the $x-y$ plane. But the world is quantum 
mechanical after all and what is obvious from the classical point of view 
often ceases to be obvious in quantum world. So we need a quantum theory of 
shapes \cite{51,51a}.

Let us consider rotations of the boron trifluoride molecule. The translational
motion is ignored as it does not lead to anything non-trivial. Therefore it
will be assumed that the center of the molecule with the boron atom at it is 
fixed at the origin. Then any configuration of the fluorine atoms can be 
obtained from a standard configuration, with one of the fluorine atoms at the 
positive y-axis, by some rotation 
$$R(\phi)=\left (\begin{tabular}{cc} $\cos{\phi}$ & $\sin{\phi}$ \\
$-\sin{\phi}$ & $\cos{\phi}$ \end{tabular} \right ). $$
But rotations by $\phi=\frac{2\pi}{3}\,n$, $n$-integer transform the molecule
into itself because of symmetry of the equilateral triangle. Therefore the
configuration space for the rotational motion of the boron trifluoride 
molecule is the coset space
$$Q=SO(2)/Z_3,$$
where $Z_3$ is the cyclic subgroup of $SO(2)$ generated by $R(2\pi/3)$.
  
Topologically SO(2) is the same as the unite circle $S^1$ and is thus
infinitely connected, because loops with different winding numbers around the
circle belong to the different homotopy classes. The configuration space $Q$ 
is obtained from $S^1$ by identifying points related by rotations with 
$\phi=\frac{2\pi}{3}\,n$, $n$-integer. Therefore $Q$ is also infinitely 
connected. The multiple connectedness brings a new flavour in quantization 
and makes it not quite trivial \cite{52,53}. For our goals, the convenient 
approach is the one presented in \cite{54,55} for quantum mechanics on the 
circle $S^1$.

Naively one expects the free dynamics of any quantum planar-rotator, such as 
the  boron trifluoride molecule, to be defined by the Hamiltonian
\begin{equation}
\hat H=\frac{\hat L^2}{2I},
\label{H} \end{equation}
where ($\hbar=1$) 
\begin{equation}
\hat L=-i\frac{\partial}{\partial \phi}
\label{Lcanon} \end{equation} 
is the angular momentum operator satisfying the canonical commutation 
relation 
\begin{equation}
[\hat \phi,\hat L]=i.
\label{phiL} \end{equation}
However, there are many pitfalls in using the commutation relation 
(\ref{phiL}) in the context of the quantum-mechanical description of angle 
variable \cite{56}. The locus of problems lies in the fact that $\hat \phi$
is not a good position operator for the configuration space $Q$, as it is 
multi-valued. Classically every point from $Q$ is uniquely determined by
a complex number $q=\exp{(-3i\phi)}$. Therefore one can expect that the 
unitary operator
\begin{equation}
\hat q=\exp{(-3i\hat\phi)},
\label{qpos} \end{equation}
is more suitable as the position operator on $Q$ \cite{57}. From (\ref{phiL})
we expect the commutation relations
\begin{equation}
[\hat L,\hat q]=-3\,\hat q,\;\;\; [\hat L,\hat q^+]=3\,\hat q^+,
\label{Lqqs} \end{equation}
which we assume to hold, although the self-adjoint angular momentum 
operator has not necessarily have to be in the naive form (\ref{Lcanon}). 

The representation of the algebra (\ref{Lqqs}) is simple to construct 
\cite{54,55}. The operator $\hat L$, being self-adjoint, has an eigenvector
$|\alpha>$ with a real eigenvalue $3\alpha$:
$$\hat L\,|\alpha>\,=3\alpha\,|\alpha>,\;\;\; <\alpha|\alpha>\,=1.$$
The commutation relations (\ref{Lqqs}) show that $\hat q$ and $\hat q^+=
\hat q^{-1}$ act as lowering and rising operators because
$$\hat L\,\hat q\,|\alpha>\,=\left ([\hat L,\hat q]+\hat q\hat L\right )
|\alpha>\,=3(\alpha-1)\hat q\,|\alpha>\;\;\mbox{and}\;\;\hat L\,\hat q^+\,
|\alpha>\,=3(\alpha+1)\hat q^+\,|\alpha>.$$
Therefore we can consider the states
$$|n+\alpha>\,=\left (\hat q^+\right )^n|\alpha>, \;\;n=0,\pm 1,\pm 2,\ldots$$
as spanning the Hilbert space ${\cal H}_{(\alpha)}$ where the fundamental 
operators $\hat L$ and $\hat q$ are realized, because, as it follows from the 
self-adjointness of $\hat L$ and unitarity of $\hat q$, the set of state 
vectors $|n+\alpha>$ forms the orthocomplete system
$$<n+\alpha|m+\alpha>\,=\delta_{nm},\;\;\sum\limits_{n=-\infty}^\infty 
|n+\alpha><n+\alpha|=1.$$ 
The angular momentum operator is diagonal in this basis
$\hat L\,|n+\alpha>\,=3(n+\alpha)\,|n+\alpha>$,
and so does the Hamiltonian (\ref{H}). The energy eigenvalues are
\begin{equation}
E_n=\frac{9(n+\alpha)^2}{2I}.
\label{En} \end{equation}
For each $\alpha$, there is a vacuum state corresponding to $n=0$ and these
vacuum states are in general different, like $\theta$-vacuums in QCD. More
precisely, ${\cal H}_{(\alpha)}$ and  ${\cal H}_{(\beta)}$ are unitary 
equivalent representation spaces of the algebra (\ref{Lqqs}) if and only if 
the difference between $\alpha$ and $\beta$ is an integer \cite{54,55}. 
Therefore, in contrast to the canonical commutation relations, the algebra 
(\ref{Lqqs}) has infinitely many inequivalent unitary representations 
parameterized by a continuous parameter $\alpha$ from the interval 
$0\le\alpha<1$.

The spectrum (\ref{En}) is doubly degenerate for $\alpha=0$, because in this 
case $E_n=E_{-n}$, as well as for $\alpha=1/2$, because then $E_n=E_{-(n+1)}$.
For other values of $\alpha$, there is no degeneracy. This degeneracy 
reflects invariance under the parity transformation \cite{55}.

As we had already mentioned, geometrically the parity transformation is the
reflection in the $y$-axis, that is inversion of $S^1$ around a diameter. 
Classically the parity transformation moves the point specified by the angle
$\phi$ to the one specified by the angle $-\phi$, if we measure an angular 
coordinate $\phi$ from the axis fixed under the parity operation. Therefore it 
is natural for the quantum mechanical unitary parity operator $\hat P$ on $Q$ 
to satisfy
\begin{equation}
\hat P^+\,\hat q\,\hat P=\hat q^+,\;\;\hat P^+\,\hat L\,\hat P=-\hat L.
\label{Pql} \end{equation}
Such a parity operator is an automorphism of the fundamental algebra 
(\ref{Lqqs}), but can not always be realized in the Hilbert space 
${\cal H}_{(\alpha)}$. Indeed,
$$\hat L\,\hat P\,|n+\alpha>\,=-\hat P\,\hat L\,|n+\alpha>\,=-3(n+\alpha)
\hat P\,|n+\alpha>$$
shows that $\hat P\,|n+\alpha>$ does not in general lies in ${\cal H}_
{(\alpha)}$, unless $\alpha=0$ or $\alpha=1/2$. Otherwise 
$\hat P\,|n+\alpha>\,\in {\cal H}_{(1-\alpha)}$. Therefore only for 
$\alpha=0$ or $\alpha=1/2$ is parity a good symmetry and other 
realizations of quantum mechanics on $Q$ break parity invariance. In 
later cases, to restore parity invariance requires doubling of the 
Hilbert space by considering ${\cal H}_{(\alpha)}\oplus{\cal H}_{(1-\alpha)}$. 

It is rather strange that most realizations of the quantum shape of the
boron trifluoride molecule violate parity, although classically the molecule
is reflection symmetric. After all no microscopic source of parity violation
was assumed in the molecular dynamics. How then does parity violation emerge
in the shape? This happens because we concentrated on the slow nuclear degrees
of freedom and completely neglected the fast electronic motion. A general 
argument is given in \cite{51,51a} that in a complete theory, with fast degrees
of freedom included, there is no parity violation. For example, in molecular
physics the coupling between rotational modes and the electronic wave 
functions lead to transitions between these wave functions related by parity,
with time scales much longer than typical rotational time scales. As a result, 
parity invariance is restored but nearly degenerate states of opposite
parity, the parity doubles, appear in the molecular spectra.  

To get an intuitive understanding why different quantum theories are possible
on $Q$ and why most of them violate parity, it is instructive to find the
explicit form for the angular momentum operator in each $\alpha$-realization
\cite{54,55}.

Let 
$$|\theta>\,=\sum\limits_{n=-\infty}^\infty a_n\,|n+\alpha>$$
be an eigenvector of the position operator $\hat q^+$:
$$\hat q^+\,|\theta>\,=e^{\,3i\theta}\,|\theta>.$$
Then we get the recurrent relation 
$$a_n=e^{-3i\theta}\,a_{n-1}$$
and, therefore, up to normalization
$$|\theta>\,=e^{i\,\omega(\theta)}\sum\limits_{n=-\infty}^\infty e^
{-3in\theta}\,|n+\alpha>,$$
where $\omega(\theta)=\omega(\theta+2\pi/3)$ is an arbitrary phase. But then
$$e^{-i\lambda\hat L}\,|\theta>\,=e^{-3i\lambda\alpha}\,e^{i\,\omega(\theta)}
\,e^{-i\omega(\theta+\lambda)}\,|\theta+\lambda>.$$
Let $|\psi>$ be an arbitrary state vector. Differentiating the equality
$$<\theta|e^{i\lambda\hat L}|\psi>=\left (<\psi|e^{-i\lambda\hat L}|\theta>
\right )^+=e^{3i\lambda\alpha}\,e^{-i\omega(\theta)}\,e^{i\,\omega(\theta+
\lambda)}<\theta+\lambda|\psi>$$
with respect to $\lambda$ and taking $\lambda=0$ in the result multiplied by 
$-i$, we get
$$<\theta|\hat L|\psi>=\left [-i\frac{\partial}{\partial\theta}+3\alpha+
\frac{\partial\omega}{\partial\theta}\right ]\psi(\theta),$$
where $\psi(\theta)=<\theta|\psi>$ is the wave function. Therefore, in the
$q$-representation, where $\hat q$ is diagonal, the angular momentum operator
is
\begin{equation}
\hat L=-i\frac{\partial}{\partial\theta}+A(\theta),
\label{alphaL} \end{equation}
with
\begin{equation}
A(\theta)=3\alpha+\frac{\partial\omega(\theta)}{\partial\theta}
\label{gaugeA} \end{equation} 
playing the role of a gauge field.

As we see, the $\alpha$-quantum theory on $Q$ is analogous to the quantum 
theory on a unit circle with the vector potential (\ref{gaugeA}) along the 
circle. Then we have the magnetic flux
$$\int \vec{B}\cdot d\vec{S}=\oint \vec{A}\cdot d\vec{l}=6\pi\alpha$$
piercing the circle perpendicularly to the $x-y$ plane. Classically a charged
particle on the circle does not feel the magnetic flux, if the the magnetic
field on the circle is zero, but quantum mechanically it does -- the 
Aharonov-Bohm effect. Therefore it is not surprising that we have many 
different quantum theories, nor is it surprising that parity is violated 
\cite{58}. It is also clear that in a more complete theory, including
sources of the magnetic flux, parity is not violated \cite{58}.  

Now back to demons. Suppose parity invariance is indeed restored in a
manner advocated by mirror matter proponents. Then some non-gravitational
interactions are not excluded between the ordinary and mirror sectors. One
interesting possibility is the photon-mirror photon kinetic mixing interaction
$${\cal{L}}=\frac{\epsilon}{2}F^{\mu\nu}F^{\,\prime}_{\mu\nu},$$
where $F^{\mu\nu}$ and $F^{\,\prime}_{\mu\nu}$ are the field strength tensors 
for electromagnetism and mirror electromagnetism respectively. As a result 
ordinary and mirror charged particles interact electromagnetically with the
interaction strength controlled by the mixing parameter $\epsilon$. A number
of observed anomalies can be explained from mirror matter perspective if
\begin{equation}
\epsilon\sim 5\cdot 10^{-9}
\label{epsilon} \end{equation} 
(see \cite{47} and references wherein).
 
Remarkably, this tiny electromagnetic interaction is nevertheless sufficient 
for thermal equilibration to occur in a mixture of ordinary and mirror matter
at Earth-like temperatures \cite{59}, and to provide a force strong enough to
oppose the force of gravity, so that a mirror matter fragment can remain on
the Earth's surface instead of falling toward its center \cite{59a}. Demons 
considered above can clearly benefit from this fact. What is necessary is to 
add a significant amount of mirror matter to the thermal reservoir we want to 
operate at colder temperature. Mirror matter will draw in heat from the 
surrounding ordinary component of the thermal bath and radiate it away as 
mirror photons. Thereby an effective cooling will take place and the necessary
temperature difference will be created between the thermal baths (assuming the 
another bath do not contain a mirror component), even if initially the 
temperatures were the same.

Equation (\ref{hflux}) indicates that, at least for some demons, heat exchange
between two thermal reservoirs of the demon can be made very low. 
Consequently, mirror matter walls of the colder reservoir should radiate a 
very small flux of mirror electromagnetic energy at dynamical equilibrium and 
hence must be very cold. In fact, it appears that for significant amount of
mirror matter the corresponding reservoir would be cooled to near absolute 
zero where approximations of reference \cite{59} break down. Therefore we
refrain from any detailed calculations. 

\section{Conclusion}
As frequently stressed by Landau \cite{39}, we expect the world of elementary
particles to be mirror symmetric because the space itself is  mirror 
symmetric. But weak interactions turned out to be left-handed and heroic 
efforts of Landau and others \cite{44,45,46} to save left-right symmetry
by {\bf CP} also failed. Therefore we are left with the strange fact that 
nature is left-right asymmetric. But the example from the molecular physics 
considered above suggests a possibility that the observed asymmetry
might be just apparent, the result of the fact that some fast degrees of 
freedom  hidden, presumably, at the Planck scale, were overlooked. Then in the
more complete theory parity symmetry will be restored, but the parity 
doubles will appear at the universe level in the form of mirror 
world \cite{38,42}.

If nature is indeed organized in this manner, the fascinating Maxwell demons
considered in the previous sections can be made operative by just adding mirror
matter to one of their thermal reservoirs. Mirror matter demons can 
extract heat from one thermal reservoir, for example from the world ocean,
and transform it very effectively (in case of Brillouin demon) to some useful
work, thereby solving the global energy problems of mankind!

All this ``sounds too good to be true'' \cite{60}. But the principal question 
is whether mirror matter exists. If it does indeed exist and if the 
photon-mirror photon mixing is not negligibly small I do not see how the 
mirror demons can fail. 

As for the photon-mirror photon mixing, the proposition that its magnitude is
of the order of (\ref{epsilon}) is experimentally falsifiable in near future,
because such mixing leads to orthopositronium to mirror orthopositronium 
oscillations and as a result to invisible decays of orthopositronium in 
vacuum, with intensities accessible in the experiment \cite{61} which is under 
way.

Maybe the putative perspective of using mirror matter demons will appear a 
little less exotic if we recall that, in a sense, we are all made of demons.
I mean that the Brownian ratchet engines, which operate extremely effectively
and reliably, can be found in every biological cell \cite{62,63,64}. Therefore
I would be not much surprised if someday nanotechnology will find the mirror 
matter useful, provided, of course, that it exists at all.

The idea that mirror matter can find applications in heat engines I 
credit to Saibal Mitra \cite{60}. But not his notes served as an inspiration 
for this investigation. The paper emerged from J.~D.~Norton's advise to be 
a little more critical about Landauer's exorcism of Maxwell's demon in 
response to my rather careless claim in \cite{65} that Landauer's principle
killed the demon. ``O King, most high and wise Lord; How incomprehensible are 
thy judgments, and inscrutable thy ways!'' 

\section*{Acknowledgments}
The author is indebted to S.~Mitra for his help with the manuscript.
Comments and suggestions from L.~B.~Okun and R.~Foot are Acknowledged with 
gratitude. Special thanks to L.~B.~Okun for constantly encouraging me to 
improve the paper. ``Have no fear of perfection - you'll never reach it'' 
-- Salvador Dali. I also have to finish without reaching it.
 
The work is supported in part by grants Sci.School-905.2006.2 and 
RFBR 06-02-16192-a.

\end{document}